# Ultra-stable Free-Space Laser Links for a Global Network of Optical Atomic Clocks

**Authors:** D. R. Gozzard[1,2,*], L. A. Howard[1], B. P. Dix-Matthews[1,2], S. F. E. Karpathakis[1], C. T. Gravestock[1], S. W. Schediwy[1,2]

[1]International Centre for Radio Astronomy Research, ICRAR M468, The University of Western Australia, 35 Stirling Hwy, Crawley, Australia 6009.

[2]Australian Research Council Centre of Excellence for Engineered Quantum Systems, Department of Physics, School of Physics, Mathematics & Computing, The University of Western Australia, 35 Stirling Hwy, Crawley, Australia 6009.

*david.gozzard@uwa.edu.au

**Abstract:** A global network of optical atomic clocks will enable unprecedented measurement precision in fields including tests of fundamental physics, dark matter searches, geodesy, and navigation. Free-space laser links through the turbulent atmosphere are needed to fully exploit this global network, by enabling comparisons to airborne and spaceborne clocks. We demonstrate frequency transfer over a 2.4 km atmospheric link with turbulence similar to that of a ground-to-space link, achieving a fractional frequency stability of $6.1\times10^{-21}$ in 300 s of integration time. We also show that clock comparison between ground and low Earth orbit will be limited by the stability of the clocks themselves after only a few seconds of integration. This significantly advances the technologies needed to realize a global timescale network of optical atomic clocks.

## I. INTRODUCTION

The establishment of a global network of optical atomic clocks will revolutionize fundamental physics experiments including tests of the General Theory of Relativity [1], investigations of the variability of fundamental constants [2], and searches for dark matter candidates [3] among others [4-6]. Other scientific and technical applications such as geodesy [7,8], satellite navigation and timing [9], and radio astronomy [10] will also benefit from such a network. To implement this network on a global scale, free-space laser links will be necessary to link clocks where optical fiber links are not available.

To fully exploit the stability and precision of the atomic clocks, frequency transfer over the free-space laser links must have residual instabilities below that of the atomic clocks comprising the network. Optical atomic clocks are now reaching instabilities as low as $2\times10^{-18}$ [11], with future clocks expected to reach this level in an integration time as short as 100 s [12]. However, atmospheric turbulence induces both phase noise on the laser signal, which degrades the phase stability of the link, and amplitude noise, which degrades the signal-to-noise ratio, or causes complete dropouts of the link due to deep fades. Phase noise is caused by time-of-flight variations due to the change in the average refractive index of the atmosphere along the path of the laser, and the phase noise is much greater than on comparable lengths of optical fiber [13-15]. Higher order turbulence modes are responsible for amplitude noise of the laser signal by causing beam wander and scintillation [14,16]. Deep fades can occur 10s to 100s of times per second for ground-to-space links [13], limiting the integration time and, thus, ultimate transfer stability and precision of the timescale comparison.

In this work, we demonstrated free-space frequency transfer with a fractional frequency stability of $6.1\times10^{-21}$ after only 300 s of integration, over a 2.4 km horizontal atmospheric link that was measured to have turbulence levels similar to a link between ground and space [17]. From this result, and taking into account the reduced bandwidth due to the greater distances involved, we predict that a ground-to-space link based on the atmospheric stabilization technology demonstrated in this paper, will achieve a fractional frequency stability of around $1\times10^{-20}$ within the viewing window of a low Earth orbit satellite transit. This means timescale comparisons via such a link will be limited by the stability of the clocks themselves after only a few seconds of integration. This achievement significantly advances the development of ground-to-space laser links for optical atomic clock comparison through the turbulent atmosphere, and provides the technological basis for the creation of a global timescale network of optical atomic clocks.

## II. EXPERIMENTAL SETUP

Ultra-stable frequency dissemination via optical fiber networks for optical atomic clock timescale comparison has been demonstrated over distances up to 1840 km [18-20] with fractional frequency instabilities as low as $6.9\times10^{-21}$ being achieved over 1400 km after an averaging time of 30,000 s [21]. Various groups [22-24] are now applying similar stabilization techniques to free-space laser links, where the amplitude noise caused by atmospheric turbulence imposes additional challenges. Notably, the Boulder Atomic Clock Optical Network Collaboration [25] recently demonstrated frequency comparison between optical atomic clocks over 1.5 km horizontal free-space link at a precision of around one part in $10^{18}$.

The technology presented here includes an optical transceiver terminal with active tip-tilt optics for amplitude stabilization, and a separate, fiberized phase stabilization system. Figure 1 shows a schematic of the experimental setup.

### A. Free-space laser link

The optical transceiver terminal was placed in a fifth-floor office in the University of Western Australia physics building, 25 m above ground level (30 m above sea level), while a 75 mm diameter corner-cube reflector (CCR) was mounted on the roof of the Harry Perkins Institute of Medical Research, 40 m above ground level (50 m above sea level), 1.2 km away, creating a 2.4 km folded horizontal free-space link as shown in Fig. 1.

### B. Tip-tilt amplitude stabilization

The amplitude stabilization system presented here is similar to that demonstrated in [22] and a simplified schematic of the system is shown in Fig. 1. The amplitude stabilization system uses a position-sensitive quadrant photodetector (QPD) to control a tip-tilt mirror to correct for beam wander and maintain alignment of the incoming light into the fiber from the phase stabilization system.

The diameter of the primary optic of the optical terminal was chosen to be 50 mm so that the diameter of the transmitted laser beam would be smaller than the Fried scale expected over the 2.4 km link [14,16] thereby requiring only first-order tip-tilt active optics for amplitude stabilization, while keeping the divergence of the laser beam as small as possible.

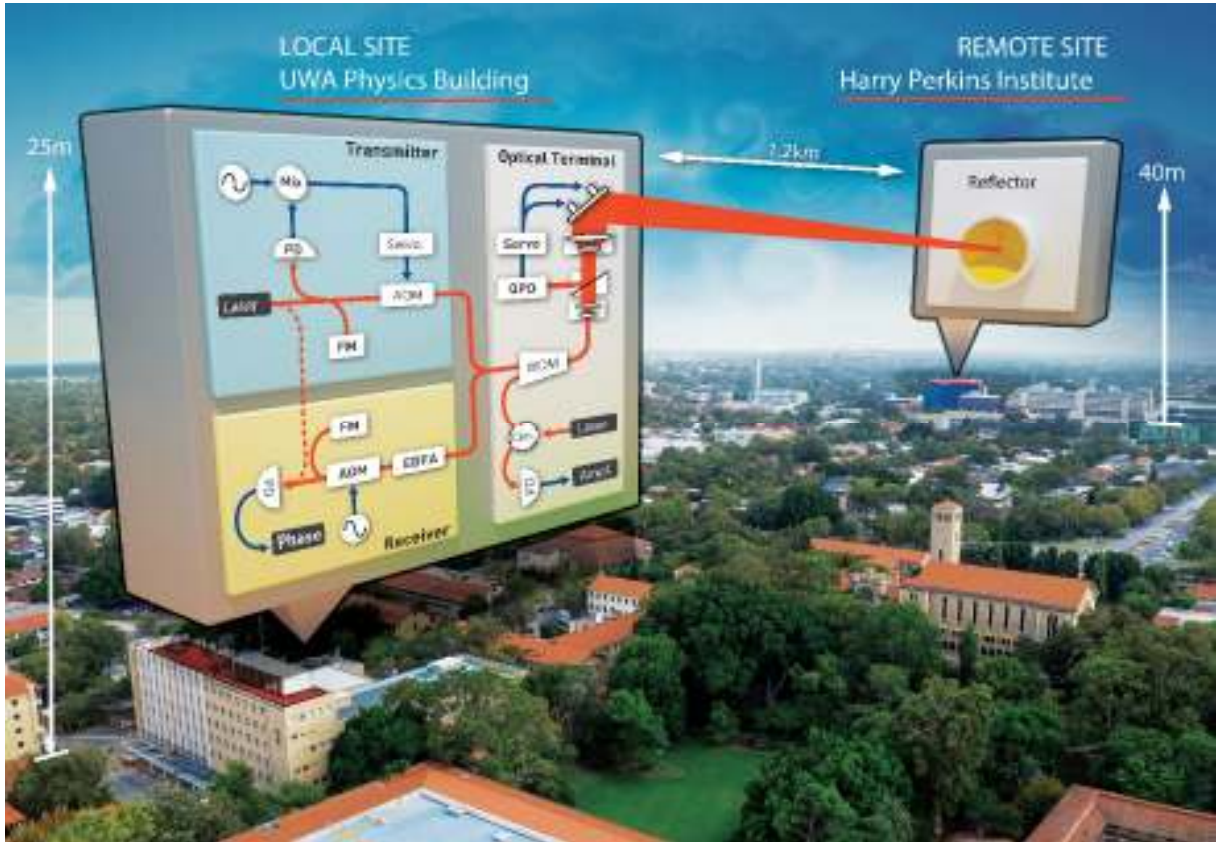

**FIG. 1.** Simplified schematic of the phase stabilization system integrated with the optical terminal. Blue lines, RF signals; and red lines, optical signals. AOM, acousto-optic modulator; Circ., circulator; FM, Faraday mirror; Mix, RF mixer; PD, photodetector; QPD, quadrant photodetector; EDFA, erbium-doped fiber amplifier; and WDM, wavelength-division multiplexer.

Light from a 20 mW (+13 dBm) C-band tunable laser, tuned to 192.7 THz (1555.75 nm), is used as a beacon for the amplitude stabilization system. This beacon light is combined with the narrow linewidth 192.7 THz (1555.75 nm) signal from the phase stabilization system (see section II.C) using a wavelength-division multiplexer (WDM), and both signals are delivered to the optical terminal via fiber. The difference in wavelength of the laser for the amplitude and phase stabilization systems is not expected to have any effect on the performance of the system, since the difference in Fried scale ($r_0$) for these wavelengths [17] is only around 0.25%. Differences between the angle of arrival of the beacon light and phase stabilization light will only be of this order.

In the terminal, the optical signals are launched into free-space via a fiber-to-free-space collimator with a beam diameter of 2.27 mm. The collimated light passes through a 50:50 cube beam splitter and the reflected light is directed to a beam dump while the transmitted light is passed to a ×15 Galilean beam expander (GBE). The GBE increases the diameter of the laser beam to 34 mm with a near diffraction limited divergence of 29 μrad. This beam is reflected off a 50 mm tip-tilt active mirror and is launched over the free-space link. The fiber-to-free-space insertion loss of the terminal, not including the 3 dB loss of the beam splitter, is around 1 dB.

The laser beam reflects off the CCR and returns to the optical terminal. The tip-tilt mirror reflects the light into the GBE and through the beam splitter. The light reflected by the beam splitter is focused by a lens onto a position-sensitive quadrant photo detector (QPD). The optical loss of the 2.4 km round trip during low turbulence due to beam divergence and scattering is 11.2 dB, not including the 3 dB loss of the beam splitter, and 7 dB free-space-to-fiber coupling loss. This means −5.2 dB (0.3 mW) of light from the tuneable laser falls on the QPD, which has a minimum operating power of −10 dB (0.1 mW).

The QPD outputs the X and Y positions of the beam centroid to a Red Pitaya field-programmable gate array (FPGA) running a pair of integral-only control loops. The output of the control loops is amplified by high-voltage amplifiers to adjust piezo actuators on the active tip-tilt mirror to keep the laser beam centered on the QPD. The control loop pauses when the returned optical power drops below a preset threshold, preventing the active mirror from railing or chasing a spurious reflection, and continues operation when sufficient optical power returns. The terminal optics are aligned to ensure that centering the beam on the QPD also maintains the optimal alignment into the fiber collimator.

To measure the amplitude fluctuations over the link, the returning beacon and phase stabilization signals are separated by the WDM and an optical fiber circulator directs the returning beacon light to a photodetector which outputs a voltage proportional to the detected optical power. This voltage was recorded via a dedicated data acquisition card at a sample rate of 20 kS/s to continuously measure the returned power over the link with sufficient time resolution to detect deep fades lasting less than a millisecond.

The optical terminal is mounted on a SkyWatcher EQ8 astronomical telescope mount which is used to provide link acquisition and slow feedback capability. The steering range of the active tip-tilt mirror is only 250 µrad, corresponding to a steering range of 30 cm at the CCR 1.2 km away, and the active mirror works most effectively over a range of only 50 µrad, meaning mechanical sagging of the mount or office floor, or thermal expansion of the building, could bring the active mirror outside of its effective steering range. This often occurred within a few hours after initial alignment of the link. The optical terminal on the mount was aligned to the CCR roughly by eye and then the mount was commanded, via a Red Pitaya, to scan the area and point to the position corresponding to highest returned optical power. After acquiring the target CCR, the tip-tilt feedback to the active mirror was activated. A separate, slow (sub-Hz), integral-only controller running on the Red Pitaya commanded the mount to move in small increments to maintain the active mirror operating in the middle of its steering range. With this slow feedback active, the link to the CCR could be maintained indefinitely.

### C. Phase stabilization

The phase stabilization system is similar to that presented in [13], and the performance characteristics and limitations of such systems have been thoroughly characterized [18]. (A thorough noise analysis for the phase stabilization system is presented in the Supplementary material [17].)

The free-space link forms the long arm of an imbalanced Michelson interferometer and the optical frequency transmitted over the link is shifted by an acousto-optic modulator (AOM). A portion of the optical signal is reflected from the phase stabilization system receiver (where it is also frequency shifted by an AOM so that unwanted reflections can be filtered out), across the link, and back to the transmitter where it creates a heterodyne beat with the short reference arm of the Michelson interferometer. Phase fluctuations on the link alter the phase of the heterodyne beat, which is mixed-down electronically with a radio-frequency reference, to generate the error signal for a phase-locked loop (PLL). This control signal adjusts the frequency of the transmitter (servo) AOM to compensate for the turbulence-induced phase noise and results in a stable optical frequency reaching the receiver.

The optics, passive electronics, and electronic amplifiers of the phase stabilization system were placed into an insulated box to minimize the influence of room temperature drifts on the stability of the stabilization system, or the out-of-loop phase measurement. Due to the optical loss through the splitter, optical terminal, and free-space link, only 1 µW of optical power reached the receiver, so a bi-directional erbium-doped fiber amplifier (EDFA) was placed at the input to the receiver to amplify the optical signal.

The performance of the phase stabilization system is measured by co-locating the transmitter and receiver, with an optical fiber splitter used to allow the system to transmit and receive via the single optical terminal. The residual phase fluctuations of the system are measured using a Microsemi 3120A phase noise test set and a Liquid Instruments Moku:Lab operating as a phasemeter. The Microsemi is able to measure phase noise up to Fourier frequencies of 100 kHz, but terminates the measurement if the signal power drops below a certain level, which can occur due to deep fading of the optical link. The Moku:Lab will continue recording after signal power has been re-established, but has a lower maximum frequency and poorer noise floor than the Microsemi. The two devices record the phase noise concurrently. The Microsemi is programmed to perform 3-minute long measurements every 15 minutes and is used to report the high-frequency performance of the system, while the Moku:Lab continuously records the phase at a sample rate of 1.95 kS/s and is used to report the long-term stability. The Microsemi and Moku:Lab results are in extremely close agreement throughout their overlapping bandwidth.

**D. Scintillometer and weather conditions**

Turbulence strength is measured using a scintillometer that was constructed in-house based on the design presented in [26]. The scintillometer comprises an 85 mm diameter, 500 mm focal length telephoto lens attached to a monochrome visible wavelength scientific camera with a global shutter operating at 200 frames per second. A green (532 nm) filter was installed in front of the camera sensor to reject unwanted light. The camera captures images of a green light-emitting diode mounted next to the CCR. The images from the camera are fed to an NVIDIA Jetson Nano which uses OpenCV to identify the centroid of the light in each image. The variance of the beam centroid is used to compute the turbulence strength, $C_n^2$, according to

$$C_n^2 = \frac{\sigma_r^2 D^{1/3}}{1.093 L f_0^2}, \quad (1)$$

where $\sigma_r^2$ is the radial variance of the centroid, $L$ is the pathlength, $D$ is the telephoto lens diameter, and $f_0$ is the focal length. The scintillometer records the measured turbulence strength every second.

In addition, weather data including wind speed and direction, temperature, humidity, and rainfall are recorded using a weather station located on the roof above the optical terminal.

**III. RESULTS**

The results were obtained over a two-week experimental campaign period in September 2020. Data acquisition runs were performed in three configurations: 1) with amplitude stabilization on and phase stabilization off, 2) amplitude stabilization off and phase stabilization on, and 3) amplitude stabilization on and phase stabilization on. Runs were conducted for up to 24 hours and stopped only to change configuration.

Figure 2 shows the fractional frequency stability of the free-space frequency transfer over the 2.4 km link expressed as modified Allan deviation (MDEV). With the phase stabilization system on, the MDEV averages down with integration time, τ, as $\tau^{-3/2}$, indicating the dominant noise source is white phase noise. This trend holds until thermal fluctuations begin to dominate the noise at integration times longer than a few tens of seconds. With the amplitude stabilization system off (blue), the best fractional frequency stability attained was $1\times10^{-19}$ at an integration time of 60 s. With the amplitude stabilization system on (orange), the fractional frequency stability improves at all integration times (down to $7\times10^{-20}$ at 60 s), as well as consistently allowing the phase synchronization to operate for significantly longer cycle-slip-free periods, reaching a best fractional frequency stability of $6.1\times10^{-21}$ after 300 s. Conveniently, this time period is similar to the typical viewing window (around 10 minutes) for a satellite in low Earth orbit. With phase stabilization off, the system attained fractional frequencies stabilities of $1.5\times10^{-13}$ at 1 s of integration, and $9.8\times10^{-15}$ at 300 s. (Additional fractional frequency results from other data runs are shown in Fig. S6 in the Supplementary material [17].)

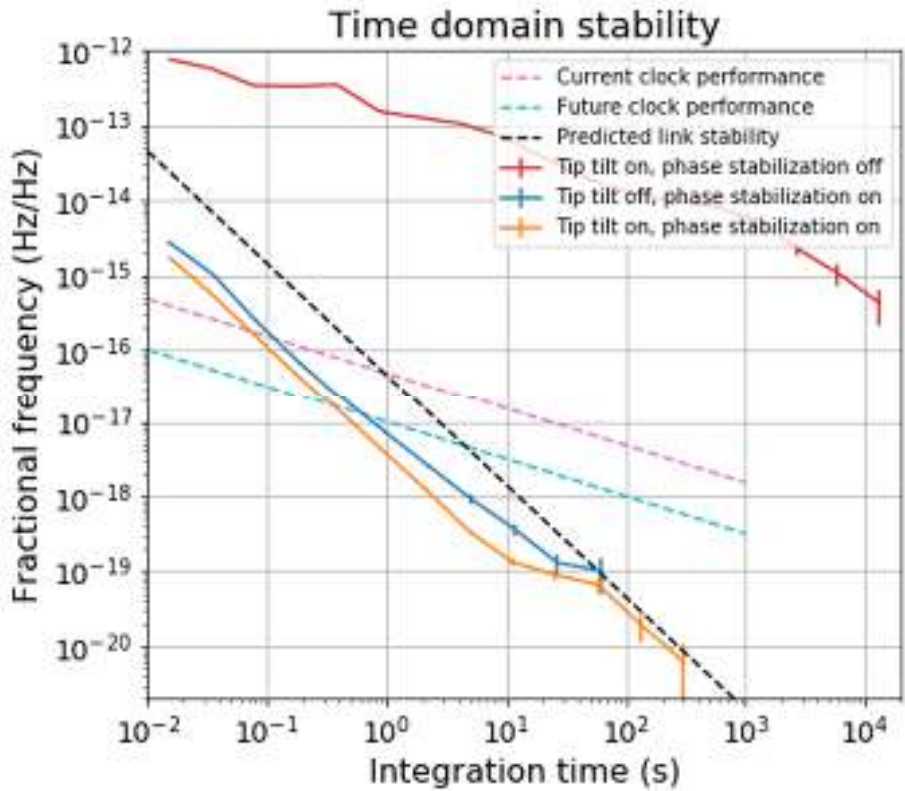

**FIG. 2.** Fractional frequency stability in terms of MDEV of the 2.4 km free-space link with amplitude stabilization on and phase stabilization off (red), amplitude stabilization off and phase stabilization on (blue), and amplitude stabilization on and phase stabilization on (orange). The error bars represent a standard fractional frequency measurement confidence interval set at $\pm\sigma_y(\tau)/\sqrt{N}$, where $\sigma_y$ is the fractional frequency stability value, $\tau$ is the integration time, and $N$ is the number of periods of $\tau$ in the data set. The dashed black line is the predicted stability for a ground-to-space link with phase stabilization bandwidth limited by the round-trip time. The dashed pink line represents current optical clock performance [11], while the dashed cyan line represents predicted performance of future optical clocks [12].

Figure 3 shows the phase noise of the frequency transfer over the 2.4 km link corresponding to the same data plotted in Fig. 2. With phase stabilization off the phase noise at 1 Hz is $2\times10^{4}$ rad$^2$/Hz, which the phase stabilization system suppresses to $2\times10^{-5}$ rad$^2$/Hz at 1 Hz when activated. The Microsemi was unable to complete any measurement runs longer than around 20 s when the amplitude stabilization system was off (including at low turbulence) due to the large power drops caused by deep fades. The measured turbulence strength for the traces in both Fig. 2 and Fig. 3 is $C_n^2 = 5\times10^{-15}$ m$^{-2/3}$.

The amplitude stabilization system significantly increased the robustness of the link and enabled regular cycle-slip-free operation for periods of around 30 minutes in low turbulence, with the longest cycle-slip free period being 104 minutes at an average turbulence strength of $C_n^2 = 6\times10^{-15}$ m$^{-2/3}$. For runs with the amplitude stabilization system off, the longest period between cycle slips was only five minutes. (A plot of the measured cycle slip rate against different turbulence strengths for amplitude stabilization on and off is shown inf Fig. S4 in the Supplementary material [17].)

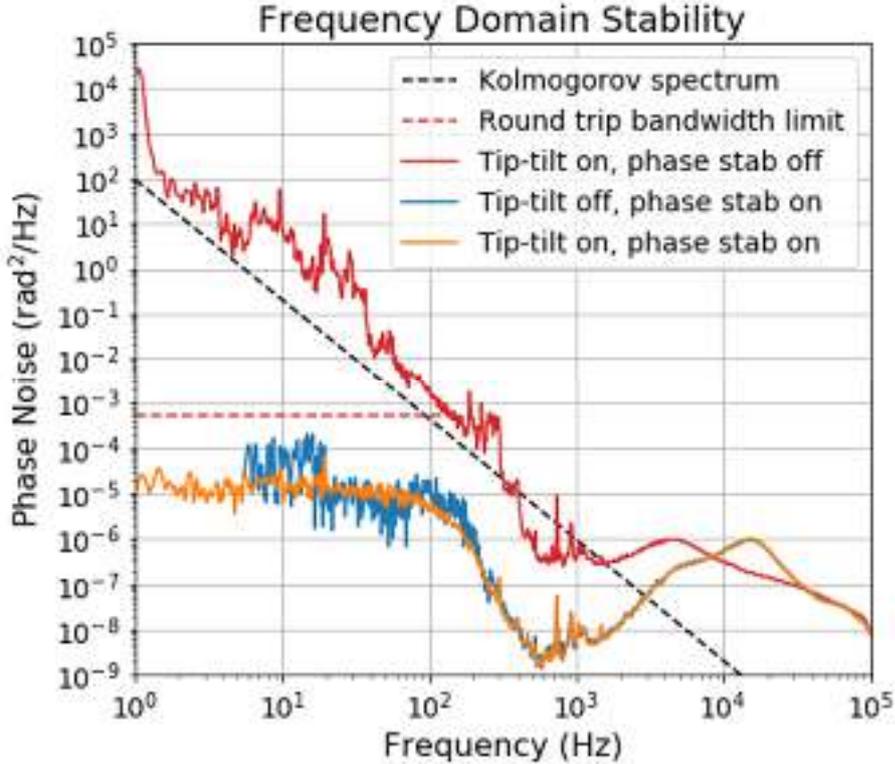

**FIG. 3.** Phase noise of the 2.4 km free-space link with amplitude stabilization on and phase stabilization off (red), amplitude stabilization off and phase stabilization on (blue), and amplitude stabilization on and phase stabilization on (orange). Dashed black line is the Kolmogorov spectrum for turbulence strength $C_n^2 = 5\times10^{-15}$ m$^{-2/3}$. Dashed red line is the compensated phase noise level achievable with a bandwidth of 150 Hz.

## IV. DISCUSSION

The phase stabilization system suppresses the phase noise of the atmosphere by nine orders of magnitude at 1 Hz. The slope of the phase noise plot is consistent with the theoretical value of $f^{-8/3}$ [27] up to a frequency of around 300 Hz. Over this range, the measured phase noise is around one order of magnitude greater than that predicted assuming Kolmogorov turbulence [15]. Measurements with the free-space link length reduced to zero meters suggest that discrepancies between theory and measurement below 500 Hz may be due to the phase noise of the length of optical fiber (3 m) between the phase stabilization system and the optical terminal.

When the phase stabilization system is on, the phase noise is flat from 1 Hz to 100 Hz before reducing at frequencies above 100 Hz. We believe this is due to broadband noise resulting from the EDFA repeatedly amplifying reflections resulting from the folded link. As a result, the fractional frequency stabilities achieved in this experiment are currently limited by the broadband noise of the optical amplifier (EDFA) at integration times shorter than 10 s. This is unlikely to be a problem in a real, bistatic, link, and this noise was not seen in point-to-point tests using the same EDFA and similar optical terminals [22].

The fractional frequency stabilities are limited by thermal instabilities at integration times longer than 10 s [17]. The optical assemblies were well insulated for passive thermal stability, and attempts to actively control the temperature did not improve the fractional frequency or phase stability. It is likely that this residual thermal instability is caused by the thermal instabilities of the measurement electronics (likely the internal phase-locked loops) and electric cables used in both the phase stabilization servo and the phase noise measurements. Future experiments with this technology will add further temperature stabilization to the control and measurement electronics in order to achieve greater longer-term stability.

Also shown in Fig. 2 are traces representing the stability of current state-of-the-art optical atomic clocks (dashed pink) [11] as well as the predicted performance of future optical clocks (dashed cyan) [12]. For a laser link to a spacecraft at an altitude of 500 km, approaching from 60° off the zenith, the light round trip time limits the phase noise correction bandwidth to approximately 150 Hz (1000 km initial distance). With this assumption, the results from Fig. 3 (backed up by Fig. S3 in the Supplementary material [17]) are used to calculate a predicted fractional frequency that could be achieved for a ground-to-space link (dashed black). (The dashed red line in Fig. 3 represents the level to which the atmospheric phase noise, solid red, could be stabilized with a bandwidth limit of 150 Hz.) This shows that, for a ground-to-space laser link using this phase and amplitude stabilization technology, measurement precision will be limited by the residual instabilities of the optical clocks after only a few seconds of integration, and will achieve a fractional frequency stability of $1\times10^{-20}$ after 300 s. This stability level is also consistent with the limits imposed by two-way differential phase noise caused by the point-ahead angle required to track the orbiting spacecraft [16,28].

Detailed analysis of the performance of the amplitude stabilization system is shown in the Supplementary material [17], including the measured cycle slip rate against different turbulence strengths (Fig. S4), and a comparison of the measured returned optical power against the expected returned optical power calculated from the measured turbulence strength (Fig. S5). The amplitude stabilization system significantly reduces the rate of cycle slips up to a turbulence strength of $C_n^2 = 1.5\times10^{-14}$ m$^{-2/3}$, corresponding to a Fried scale of 36 mm over the 2.4 km link, very close to the fiber-coupled beam diameter of the optical terminal (see section 5 in the Supplementary material [17]). In conventional adaptive optics literature [29], it is often stated that first-order tip-tilt active optics should effectively suppress spatial distortions caused by the atmosphere until the diameter of the aperture is two to three times larger than the Fried scale. However, the rapid reduction in performance of the amplitude stabilization system when the Fried scale becomes comparable to, or smaller than, the fiber-coupled beam diameter suggests that higher-order adaptive optics are needed at much lower turbulence levels. This is consistent with previously reported results using similar optical terminals [16]. This may not be a problem for ground-to-space links, where the Fried parameter will be on the order of 10 cm or greater, but horizontal ground-to-ground links will either have to employ higher-order adaptive optics, or use a smaller aperture and suffer the resulting additional signal loss due to increased beam divergence.

## V. CONCLUSION

The phase and amplitude stabilization technologies described in this paper are shown to be able to provide a robust and stable laser link over a 2.4 km horizontal free-space path with levels of turbulence similar to that expected over a slant-path link to a low Earth orbit satellite. The tip-tilt active optics are able to effectively suppress the amplitude noise caused by beam wander until the Fried scale becomes comparable in size to the primary optic of the optical terminal, after which point, higher-order adaptive optics are necessary. The phase stabilization system is able to stabilize the extremely high levels of phase noise caused by atmospheric turbulence and, with the improved signal-to-noise ratio and longer cycle-slip-free periods made possible by the tip-tilt stabilization, is able to achieve a fractional frequency stability of $6.1 \times 10^{-21}$ after 300 s of integration, a time period comparable to the viewing window of a satellite in low Earth orbit. Over the 2.4 km link, the residual instability of the phase stabilization system is better than that of modern optical atomic clocks [11] after less than 100 ms of integration.

The measured unstabilized phase noise of the 2.4 km horizontal link is greater than that expected for a 1000 km slant-path to a low Earth orbit satellite 60° off the zenith. By limiting the phase noise suppression bandwidth of the phase stabilization system to 150 Hz, limited by the light round-trip time to the satellite, we are able to estimate the achievable fractional frequency stability of a laser link from the ground to low Earth orbit. We find that such a link would have a fractional frequency stability of around $4.5 \times 10^{-17}$ after 1 s of integration. This is consistent with the expected stability limit due to the differential path noise caused by the point-ahead angle to the satellite [14,28], and means that a ground-to-space link using this stabilization technology will surpass the residual instabilities of modern optical clocks after only around 1 s of integration.

This result shows that the phase and amplitude stabilization technologies presented in this paper can provide the basis for ultra-precise timescale comparison of optical atomic clocks through the turbulent atmosphere. We are currently deploying upgraded versions of these systems over a 10 km horizontal link to investigate more challenging atmospheric turbulence scenarios, and are also developing ground terminals able to track drones, airplanes, and high-altitude balloons as precursors to establishing a link to a satellite in orbit. Tracking moving targets requires Doppler compensation, and high-stability frequency-swept laser systems [30] are also being developed to account for the large and rapidly changing Doppler shifts necessary to maintain an optical metrology link to a satellite in low Earth orbit. Applying these technologies to advanced optical ground stations currently under development [31], will enable the creation of a global optical atomic clock network for high-precision fundamental and applied science measurements limited by the residual instabilities of the optical atomic clocks. This technology also has promising applications in free-space laser communications [32] and quantum key distribution [33].

## ACKNOWLEDGMENTS

The authors would like to thank the Harry Perkins Institute of Medical Research for assistance in setting up the free-space link. This research is funded by the Australian Research Council's Centre of Excellence for Engineered Quantum Systems (EQUS, CE170100009) and the SmartSat Cooperative Research Centre (CRC. Research Project 1-01). D.R.G. is supported by a Forrest Research Foundation Fellowship. B.P.D-M. is supported by an Australian Government Research Training Program (RTP) Scholarship and a CRC Top-Up Scholarship.

D.R.G. and L.A.H. contributed equally to this work.

# Supplementary Material
# Ultra-stable Free-Space Laser Links for a Global Network of Optical Atomic Clocks

**Authors:** D. R. Gozzard, L. A. Howard, B. P. Dix-Matthews, S. F. E. Karpathakis, C. T. Gravestock, S. W. Schediwy

1. Phase stabilization system noise analysis

Here we estimate the noise floor of the phase stabilization system resulting from residual atmospheric noise and laser frequency noise when the phase stabilization system is active. For a stationary noise process x(t) that has a PSD Sx(f), the linear combination of noise processes is given by

$$S_z(f) = \left(\sum_i a_i^2 + \sum_i a_i a_j \cos\left(2\pi f(T_i - T_j)\right)\right) S_x(f), \tag{1}$$

where $a_i$ and $a_j$ are arbitrary real constants.

Assuming a link delay of $T$, the heterodyne beat, $\nu_{servo}$, at the phase stabilization system servo photodetector is:

$$\nu_{servo} = -\nu_c(t) + \nu_c(t - 2T) + \nu_{atm}(t - T) + \nu_{atm}(t) + \nu_{AOM}(t) + \nu_{AOM}(t - 2T), \tag{2}$$

where $\nu_c$ is the optical carrier frequency, $\nu_{atm}$ is the noise of the atmosphere, and $\nu_{AOM}$ is the frequency shift of the servo AOM.

Under the condition that the PLL drives the servo signal to zero and $\nu_{AOM}(t) \approx \nu_{AOM}(t - 2T)$, $\nu_{AOM}$ becomes

$$\nu_{AOM} = \frac{1}{2}[\nu_c(t) - \nu_c(t - 2T) - \nu_{atm}(t - T) - \nu_{atm}(t)], \tag{3}$$

The beat frequency at the phase noise measurement photodetector is given by

$$\nu_{PN} = -\nu_c(t) + \nu_c(t - T) + \nu_{atm}(t) + \nu_{AOM}(t - T), \tag{4}$$

and, using the equation for $\nu_{AOM}$, becomes

$$\nu_{PN} = -\nu_c(t) + \frac{3}{2}\nu_c(t - T) - \frac{1}{2}\nu_c(t - 3T) + \nu_{atm}(t) - \frac{1}{2}\nu_{atm}(t - T) - \frac{1}{2}\nu_{atm}(t - 2T). \tag{5}$$

Applying the formula for $S_z(f)$, the contributions of the residual laser and atmospheric noise sources at the phase noise measurement photodetector are respectively

$$S_{meas}^c(f) = \left(\frac{7}{2} - 3\cos(2\pi fT) - \frac{3}{2}\cos(4\pi fT) + \cos(6\pi fT)\right) S_c(f), \tag{6}$$

And

$$S_{meas}^{atm}(f) = \left(\frac{3}{2} - \frac{1}{2}\cos(4\pi fT) - \cos(4\pi fT)\right) S_{atm}(f), \tag{7}$$

The propagation time, $T$, is estimated from the optical paths lengths, which are 5 m of optical fiber internal to the phase stabilization system, 5 m of optical fiber patch cord from the phase stabilization system to the optical terminal, and the 2.4 km free-space link. The index of refraction of the optical fiber is 1.45. Figure S1 shows the unstabilized atmospheric noise (red, from Fig. 3), the residual laser phase noise (green), and the theoretical stabilized phase noise performance of the phase stabilization system over the 2.4 km link (purple).

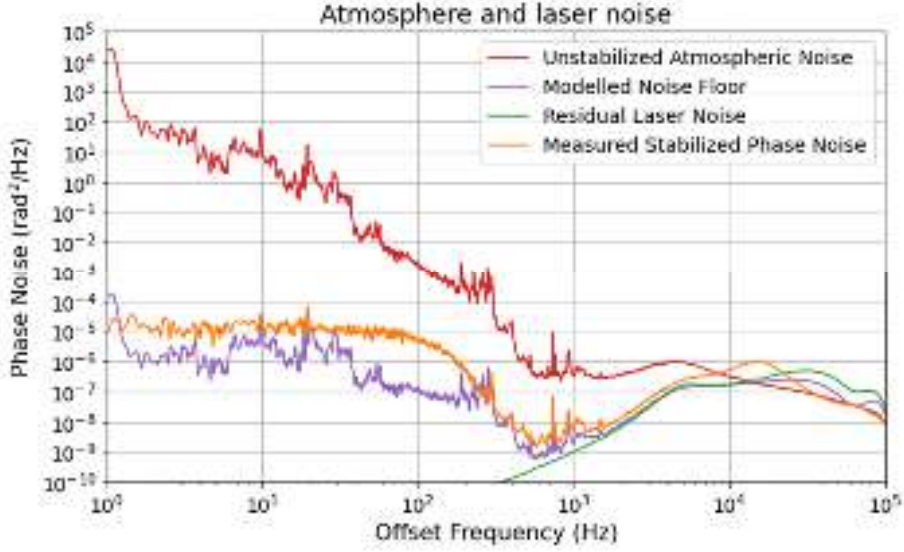

**Fig. S1.**

Estimate of the noise floor of the phase stabilized 2.4 km free-space link. Red, unstabilized atmospheric noise from Fig. 3; green, residual noise of laser; purple, calculated phase noise floor; orange, measured phase noise performance from Fig. 3.

The measured phase noise (orange) is consistent with the modelled phase noise floor at frequencies above 20 Hz. Measurements over shorter and point-to-point links [22] indicate that the excess phase noise seen in the measured phase noise performance is mainly attributed to the EDFA.

2. Turbulence effects

The Fried coherence length, $r_0$, is the diameter of a circular aperture over which the RMS wavefront aberration due to atmospheric turbulence is 1 rad. This coherence length, or 'Fried scale', defines the maximum diameter of an optical receiver aperture before atmospheric distortion begins to significantly impact the performance. Under the assumption of a horizontal atmospheric link with uniform turbulence distribution along the path (i.e. $C_n^2$ is constant along the path), the Fried coherence length for a plane wave reduces to [29]

$$r_0 = 1.68(C_n^2 L k^2)^{-3/5}, \qquad (8)$$

where $L$ is the length of the path and $k$ is the wave number of the optical signal.

Under the assumption of Taylor's frozen flow hypothesis and Kolmogorov turbulence, the phase noise spectrum, $S_\varphi(f)$, caused by atmospheric turbulence is given by [15]

$$S_\varphi(f) = 0.016 k^2 C_n^2 L V^{5/3} f^{-8/3} \quad [rad^2/Hz], \qquad (9)$$

where $V$ is the windspeed perpendicular to the beam direction and $f$ is the Fourier frequency of the turbulence. This was used to compute the Kolmogorov turbulence spectrum shown in Fig. 3. The phase noise spectrum can be converted to the angle of arrival spectrum, $S_\alpha(f)$, by

$$S_\alpha(f) = \left(\frac{c}{v_c V}\right)^2 f^2 S_\varphi(f) \quad [rad^2/Hz], \qquad (10)$$

where $v_c$ is the optical carrier frequency and $c$ is the speed of light.

## 3. Comparison to ground-to-space link

We will assume the ground-to-space link is targeted at a satellite in a circular low Earth orbit at an altitude of 500 km and an orbital speed of 7.68 km/s. Using the Hufnagel-Valley model, the $C_n^2$ profile with altitude is [29]

$$C_n^2(h) = 0.00594(w/27)^2(10^{-5}h)^{10}\exp\left(-\frac{h}{100}\right) + 2.7\times10^{16}\exp\left(-\frac{5}{1500}\right) + A\exp\left(-\frac{h}{100}\right), \quad (11)$$

where $h$ is the height above the ground, $A$ is the value of $C_n^2$ at ground level, and $w$ is the 'pseudowind' given by

$$w = \sqrt{\frac{1}{15\times10^3}\int_{5\times10^3}^{20\times10^3} V^2(h)dh}, \quad (12)$$

where $V(h)$ is given by the Bufton wind model as

$$V(h) = w_s h + V_g + 30\exp\left[-\left(\frac{h-9400}{4800}\right)^2\right], \quad (13)$$

where ws is the slew rate of the ground station and $V_g$ is the ground wind speed (assumed to be 10 km/h for this analysis).

Figure S.2 shows the angle of arrival spectrum for the horizontal 2.4 km link and for a ground-to-space link computed from the above models. The calculations assume a windspeed of 10 km/h, a turbulence strength of $C_n^2 = 1\times10^{-14}$ m$^{-2/3}$ at ground level, and that the ground terminal is pointing at a spacecraft 60° off the zenith (i.e. approximately 1000 km from the ground station).

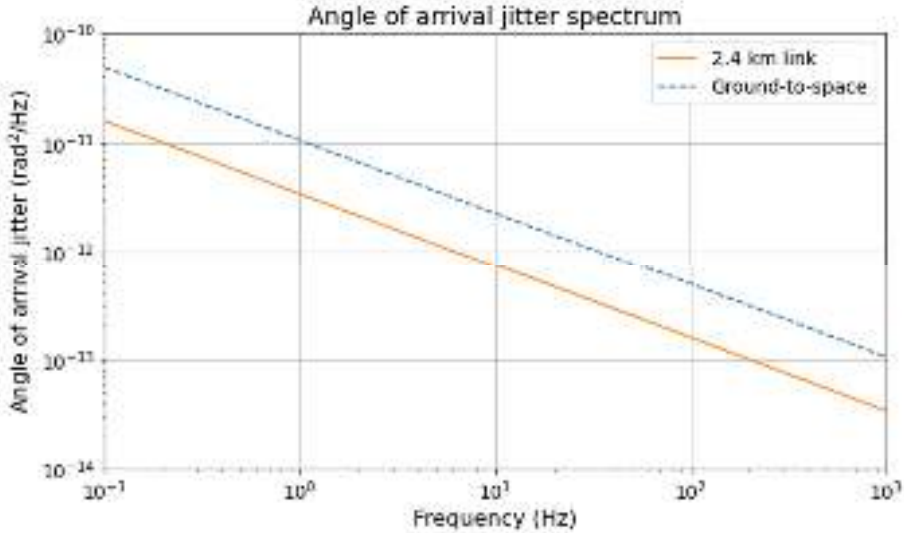

**Fig. S2.**

Laser beam angle of arrival jitter spectrum for the 2.4 km horizontal link (orange) and ground-to-space link (dashed blue).

Figure S.3 shows the theoretical phase noise spectrum for the 2.4 km link and for a ground-to-space link with the same assumptions as for the angle of arrival jitter spectrum. The figure also shows the measured phase noise over the 2.4 km link for comparison.

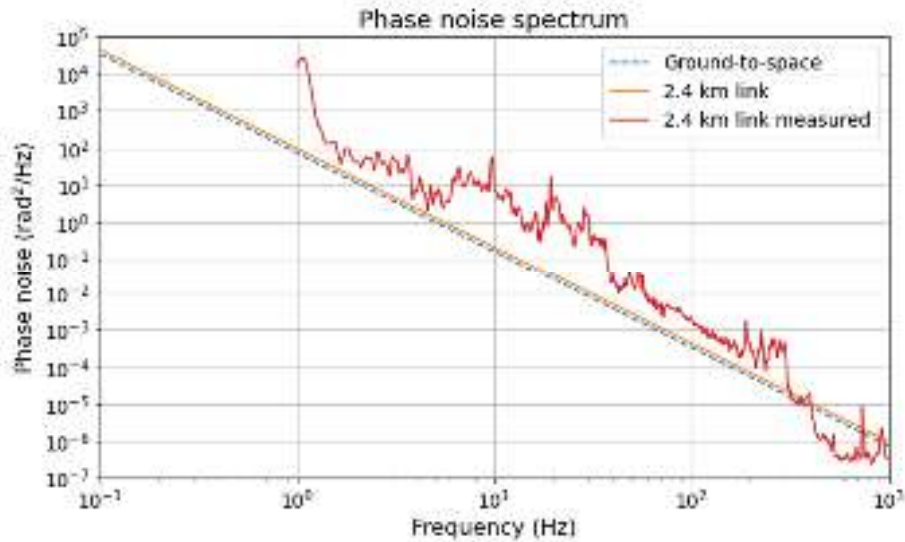

**Fig. S3.**

Expected phase noise spectrum for 2.4 km link (orange) and ground-to-space link (dashed blue). Measured 2.4 km link phase noise (red) shown for comparison.

It can be seen that the angle of arrival jitter spectrum for the 2.4 km horizontal link is of the same order of magnitude as for a slant path sweeping through the atmosphere to a LEO spacecraft for the same turbulence strength at ground-level. For these scenarios, the calculated phase noise spectrum is very similar, while the actual measured phase noise over the 2.4 km link is significantly greater than the predicted level. Therefore, we can conclude that the turbulence effects over the 2.4 km horizontal free-space link are similar to or greater than a ground-to-space link from the same location in the same conditions.

4. Amplitude stabilization system performance

Figure S4 shows the rate of cycle slips against atmospheric turbulence strength with amplitude stabilization off (blue) and on (orange).

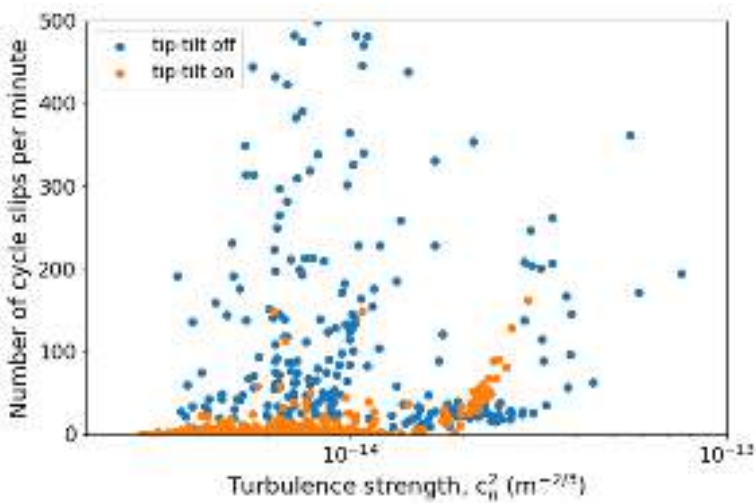

**Fig. S4.**

Rate of cycles slips of phase against atmospheric turbulence strength for spatial stabilization off (blue) and on (orange).

5. Optical power vs turbulence

The power, P, received through a circular aperture is given by
$$P = P_0[1 - \exp(-2r_A^2/w_z^2)], \quad (14)$$

where $P_0$ is the power of the incident beam, $r_A$ is the radius of the aperture, and $w_z$ is the radius of the beam arriving at the aperture. Because, in these experiments, the returning laser beam is coupled back into fiber, only the central coherence region of the beam will couple efficiently into the single mode fiber. Thus, in the case where turbulence is sufficiently strong that $r_0$ is smaller than $r_A$, $r_0$ is used in place of $r_A$ to calculate the received power.

To calculate $w_z$ it is necessary to take into account the additional beam divergence caused by atmospheric turbulence. The beam divergence, $\theta_{turb}$, taking atmospheric turbulence into account, it given by
$$\theta_{turb} = \frac{1}{2}\sqrt{\left(\frac{4\lambda}{\pi r_A}\right)^2 + \left(\frac{2\lambda}{r_0}\right)^2}, \quad (15)$$

where λ is the wavelength of the optical carrier.

Accounting for free-space-to-fiber coupling loss, loss through the terminal optics, and the efficiency and gain of the photodetector, it is possible to calculate the expected optical power (photodetector voltage) for a measured $C_n^2$. Figure S5 shows the measured turbulence strength (orange), expected photodetector voltage given the measured turbulence strength (light blue), and the measured photodetector voltage (dark blue).

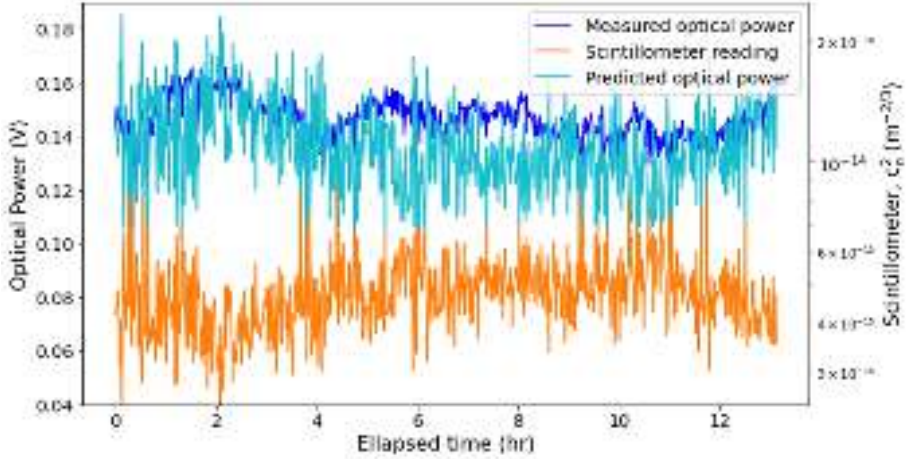

**Fig. S5.**
Measured returned optical power, in terms of photodetector voltage (blue) compared with expected optical power (orange) calculated from measured turbulence strength (green).

It can be seen from Fig. S5 that the expected voltage tracks the measured voltage very well during some periods, and less well during others. We surmise that periods when the expected voltage does not track the measured voltage are due to turbulence not being uniformly distributed across the free-space path during those periods. For example, heating of, and air movement around, the buildings at the ends of the link will cause a non-uniform turbulence distribution. No significant correlation of turbulence strength, or degree of correlation between expected and measured voltage, was found to either local wind velocity or air temperature.

6. Additional fractional frequency stability results

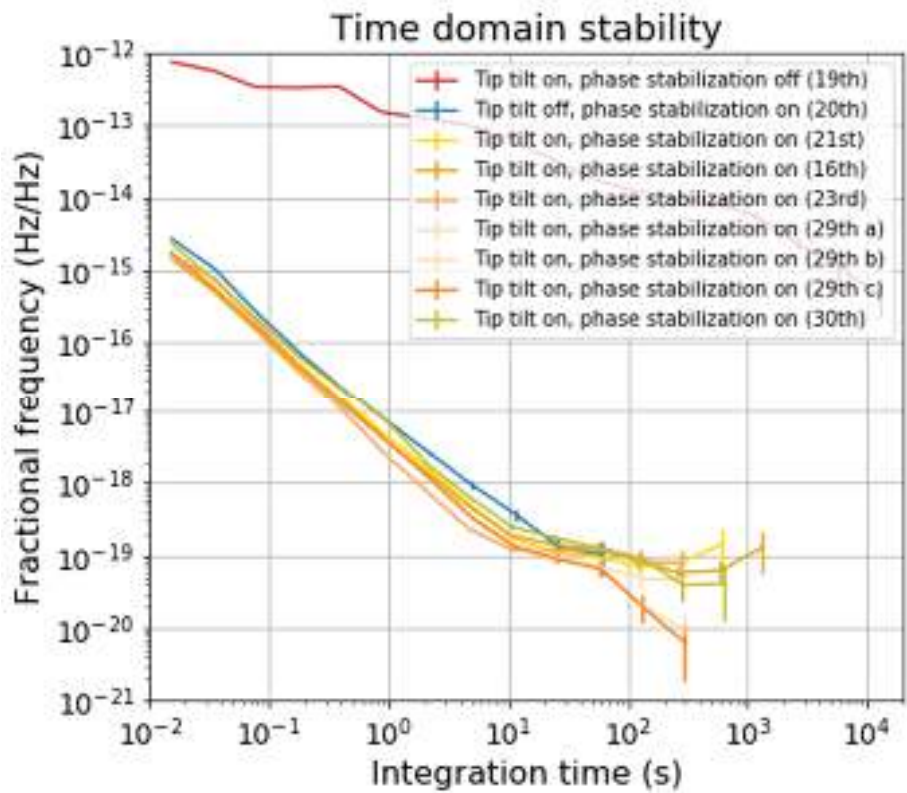

**Fig. S6.**
Figure 2 with additional fractional frequency traces to show repeatability of long-term stability performance. Dates in brackets indicate the day in September 2020 this data is from. The error bars represent a standard fractional frequency measurement confidence interval set at $\pm\sigma_y(\tau)/\sqrt{N}$, where $\sigma_y$ is the fractional frequency stability value, $\tau$ is the integration time, and $N$ is the number of periods of $\tau$ in the data set.